\documentclass[%
 aip,
 amsmath,amssymb,
 reprint,%
]{revtex4-1}

\usepackage{graphicx}
\usepackage{dcolumn}
\usepackage{bm}

\usepackage[utf8]{inputenc}
\usepackage[T1]{fontenc}
\usepackage{mathptmx}
\usepackage{soul}

\begin{document}

\preprint{AIP/123-QED}

\title{Voltage control of skyrmions: creation, annihilation and zero-magnetic field stablization}

\author{Yifan Zhou}
\affiliation{NanoSpin, Department of Applied Physics, Aalto University School of Science, P.O. Box 15100, FI-00076 Aalto, Finland}
\author{Rhodri Mansell}
\email{rhodri.mansell@aalto.fi}
\affiliation{NanoSpin, Department of Applied Physics, Aalto University School of Science, P.O. Box 15100, FI-00076 Aalto, Finland}
\author{Sebastiaan van Dijken}
\affiliation{NanoSpin, Department of Applied Physics, Aalto University School of Science, P.O. Box 15100, FI-00076 Aalto, Finland}

\date{\today}

\begin{abstract}
Voltage manipulation of skyrmions is a promising path towards low-energy spintronic devices. Here, voltage effects on skyrmions in a GdO$_{x}$/Gd/Co/Pt heterostructure are observed experimentally. The results show that the skyrmion density can be both enhanced and depleted by the application of an electric field, along with the ability, at certain magnetic fields to completely switch the skyrmion state on and off. Further, a zero magnetic field skyrmion state can be stablized under a negative bias voltage using a defined voltage and magnetic field sequence. The voltage effects measured here occur on a few-second timescale, suggesting an origin in voltage-controlled magnetic anisotropy rather than ionic effects. By investigating the skyrmion nucleation rate as a function of temperature, we extract the energy barrier to skyrmion nucleation in our sample. Further, micromagnetic simulations are used to explore the effect of changing the anisotropy and Dzyaloshinskii-Moriya interaction on skyrmion density. Our work demonstrates the control of skyrmions by voltages, showing functionalities desirable for commercial devices.
\end{abstract}

\maketitle

Magnetic skyrmions are topologically non-trivial spin textures, which are widely observed in thin film magnetic trilayers consisting of a heavy metal (HM), a ferromagnet (FM) and a metal oxide (MO), such as Pt/Co/MgO\cite{boulle2016room}, Pt/CoFeB/MgO\cite{woo2016observation}, Ta/CoFeB/MgO\cite{gilbert2015realization}, Ta/CoFeB/TaO$_x$\cite{jiang2015blowing} and so on\cite{everschor2018perspective}. A combination of perpendicular magnetic anisotropy (PMA) and the Dzyaloshinskii–Moriya interaction (DMI) can lead to a skyrmion state\cite{roessler2006spontaneous,nagaosa2013topological,fert2013skyrmions} in such systems. The strong spin-orbit coupling at the HM/FM interface gives rise to PMA\cite{hellman2017interface} and, in combination with the broken inversion symmetry in the growth direction,  DMI\cite{yang2015anatomy}. Further, PMA and DMI originate from the FM/MO interface due to the overlapping $p$ orbitals from oxygen and $d$ orbitals from the ferromagnet\cite{yang2011first}.  Skyrmions have attractive features for device applications, such as small sizes\cite{wang2018theory}, stability at room temperature\cite{buttner2018theory}, and can be driven into motion by a relatively low current density\cite{iwasaki2013current}. However, the skyrmion Hall effect\cite{jiang2017direct}, as well as the Joule heating produced by driving currents\cite{koshibae2015memory,koshibae2014creation}, has hindered the application of skyrmions to current-driven memory devices.

As an alternative approach to current-driven devices, the voltage control of magnetism has been widely investigated, initially in magnetic tunnel junctions\cite{wang2012electric,kanai2012electric,shiota2012induction} and then other FM/MO systems\cite{bi2014reversible,bauer2015magneto,baldrati2017magneto}. Several proposals have been made for skyrmion-based voltage controlled memory devices, using both static\cite{Bhattacharya2016VCMASwitch,Kasai2019SkMTJ} and mobile\cite{Wang2018VCMARacetrack,Zhou2019Gyro} skyrmions. In HM/FM/MO structures hosting skyrmions, recent experiments have demonstrated the voltage control of magnetic anisotropy (VCMA), with the additional ability to control DMI by applying voltages\cite{srivastava2018large,yang2020voltage}. In such experiments, due to the modification of these underlying magnetic properties, skyrmions can be created and annihilated by applied voltages. Different mechanisms have been proposed which would allow the voltage control of skyrmions, namely changing the electron orbital filling with an electric field\cite{hsu2017electric,schott2017skyrmion,bhattacharya2020creation}, modifying the Rashba DMI field at FM/MO interface\cite{srivastava2018large}, and introducing strains from flexible \cite{yang2020voltage} or ferroelectric \cite{Li2018Strain,Wang2020FEFMSk} substrates. Inspired by the control of magnetism through ionic effects demonstrated in Pt/Co/GdOx heterostructures\cite{tan2019magneto,bauer2015magneto}, we explore the possibility of observing skyrmions in such a structure, and subsequently controlling them by applied voltages.

\begin{figure}[!h]
\centering
\includegraphics[width=\columnwidth]{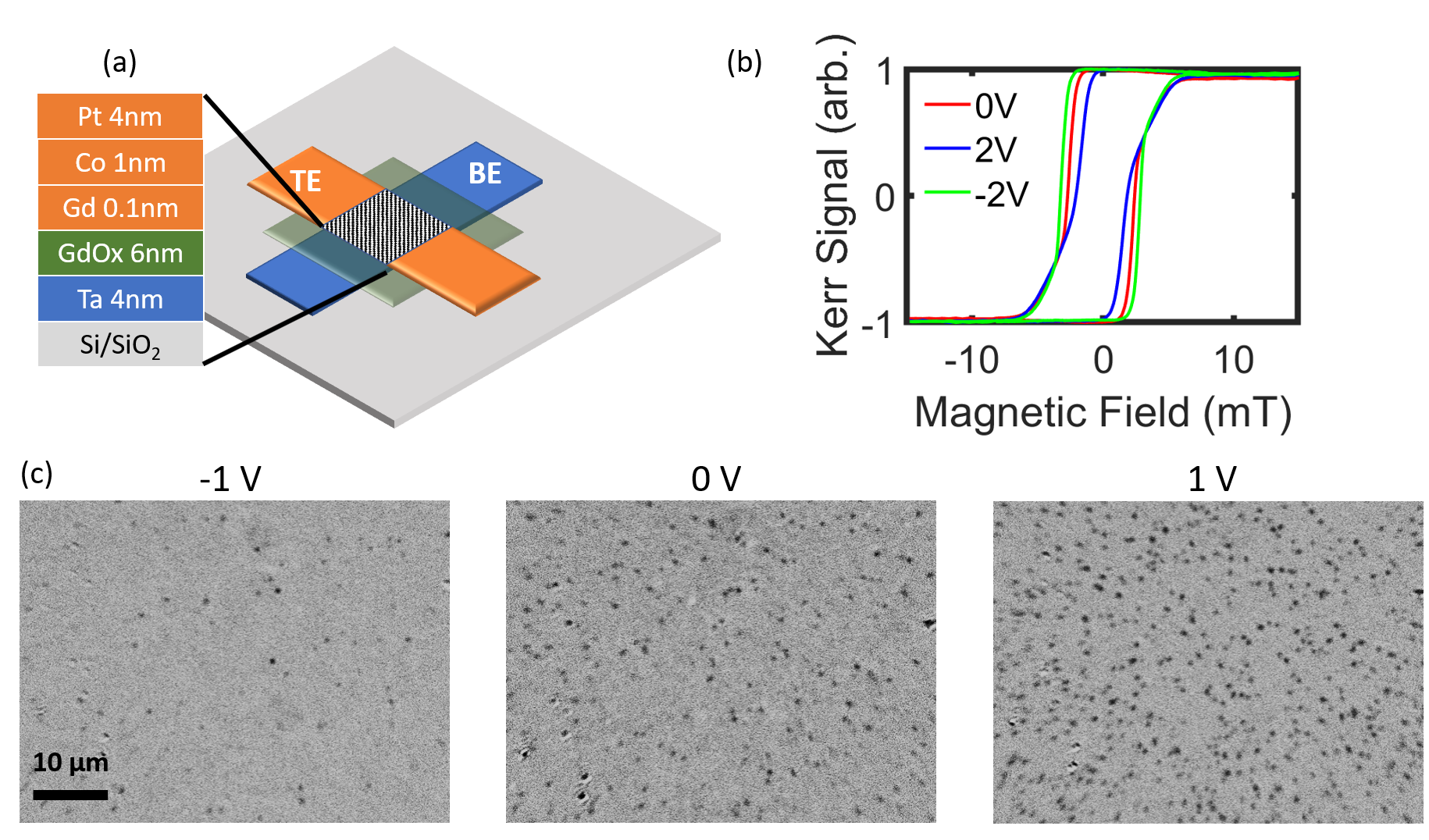}
\caption{(a) Schematic of the multilayer sample with a cross bar structure. (b) Hysteresis loops obtained by MOKE microscopy with constant bias voltages of 0 V, 2 V and -2 V. (c) Skyrmion states at 3 mT with different bias voltages of -1 V, 0 V and 1 V.}
\label{fig1}
\end{figure}
The thin film sample studied in this work is a Ta(4) / GdO$_x$(6) / Gd(0.1) / Co(1) / Pt(4) (in nm) heterostructure. The sample is deposited by magnetron sputtering at room temperature in a system with a base pressure of $\sim 5 \times 10^{-8} $ mbar. Metal layers are grown by DC sputtering with an Ar pressure of $8\times 10^{-3}$ mbar, while the GdO$_x$ layer is grown by reactive DC sputtering with 10\%  O$_2$ partial pressure. The sample is grown with an `inverted' layer structure, with the magnetic metal layer grown on top of the GdO$_x$. The introduction of the thin Gd metal layer acts to reduce the oxidation of the Co layer. As shown in Fig.\ 1(a), the multilayer is patterned into a crossbar structure by direct laser-writing lithography. In the patterned junction, Ta is the bottom electrode (BE), GdO$_x$ is an insulating layer and the Gd/Co/Pt multilayer is the top electrode (TE). The junction area is 50 $\mu$m $\times$ 50 $\mu$m.  The sign of the applied voltage is defined from the top electrode to the bottom electrode, where a positive sign means the voltage on the top electrode is higher than on the bottom electrode. 

Magneto-optical Kerr effect (MOKE) microscopy is used to record and image the out-of-plane magnetization of the sample, under externally applied out-of-plane magnetic fields and voltages. Magnetic properties of the sample with zero voltage, such as the saturation magnetisation $M_s$ and out-of-plane anisotropy $K_u$, were measured on an equivalent thin film sample by vibrating sample magnetometry (VSM) at $25^{\circ}$C, where $M_s = 1.4 \times 10^6$ A/m and $K_u=6\times 10^6$ J/m$^3$, consistent with previously reported values from a similar structure\cite{pham2016very}.  

To study voltage effects on the magnetic properties of the sample, we first measure the out-of-plane hysteresis loop 
with a constantly applied voltage of 0 V, 2 V and -2 V using MOKE microscopy (Fig.\ 1(b)). The results show a perpendicularly magnetized sample with near to full remanence at zero applied voltage. A positive bias voltage decreases the coercivity of the sample, indicating a reduction of the perpendicular anisotropy, or possibly an increase in DMI, while a negative bias has the reverse effect. The voltage effect is volatile, meaning that after any applied voltage is removed, the 0 V hysteresis loop is the same as before the application of voltage. Images of skyrmion states are captured at different bias voltages by first saturating the sample at 10 mT and then decreasing the field to $3$ mT (Fig.\ 1(c)). The images are taken after a relaxation time of one minute to allow for skyrmion nucleation. The voltage is applied throughout this process. In spite of the relatively small voltage effect seen in the hysteresis loops, the skyrmion density varies significantly with bias voltage, where more skyrmions are observed with positive voltage and fewer with negative voltage. Due to the resolution limit ($\sim 500$ nm) of white-light MOKE microscopy, we are not able to observe variations of the skyrmion radius, which might be expected from voltage-induced changes of the magnetic anisotropy.     
\begin{figure}[!h]
\centering
\includegraphics[width=\columnwidth]{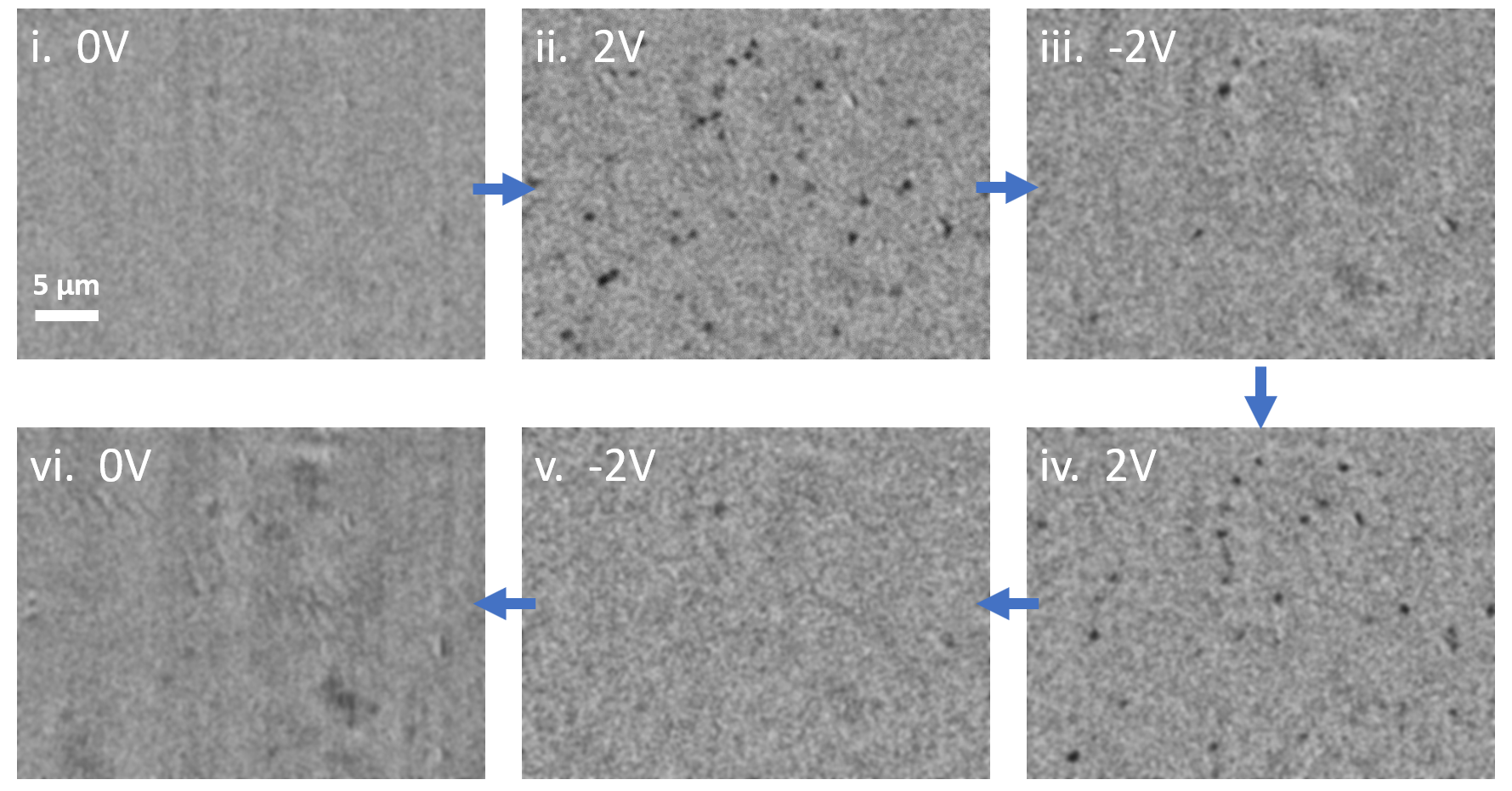}
\caption{Real time control of skyrmion creation and annihilation from a uniform magnetization state at 3.5 mT with a voltage sequence of i. 0 V, ii. 2 V, iii. -2 V, iv. 2 V, v. -2 V,  vi. 0 V. Each image is taken 30 s after changing the applied voltage.}
\label{fig2}
\end{figure}

On-off control of skyrmions starting from a uniform magnetization state can be  achieved by applying a suitable voltage sequence (Fig. 2). The sample is initially saturated by a 3.5 mT out-of-plane field, which is slightly larger than the transition field from an uniform state to the skyrmion state at $3$ mT (Fig.\ 2i.). 
After this, a 2 V bias is applied continuously while the magnetic field is fixed at 3.5 mT, and the sample is imaged by MOKE microscopy after 30 s (Fig.\ 2ii.). Under these conditions, skyrmions are created by the positive bias voltage. 
The voltage is then set to -2 V (Fig.\ 2iii.), and after 30 s most of the skyrmions have disappeared.  
Repetition of the same voltage sequence (Fig.\ 2iv. and 2v.) produces a similar effect. The system returns to a uniform magnetization state at 0 V (Fig.\ 2vi.). 
\begin{figure}[!h]
\centering
\includegraphics[width=\columnwidth]{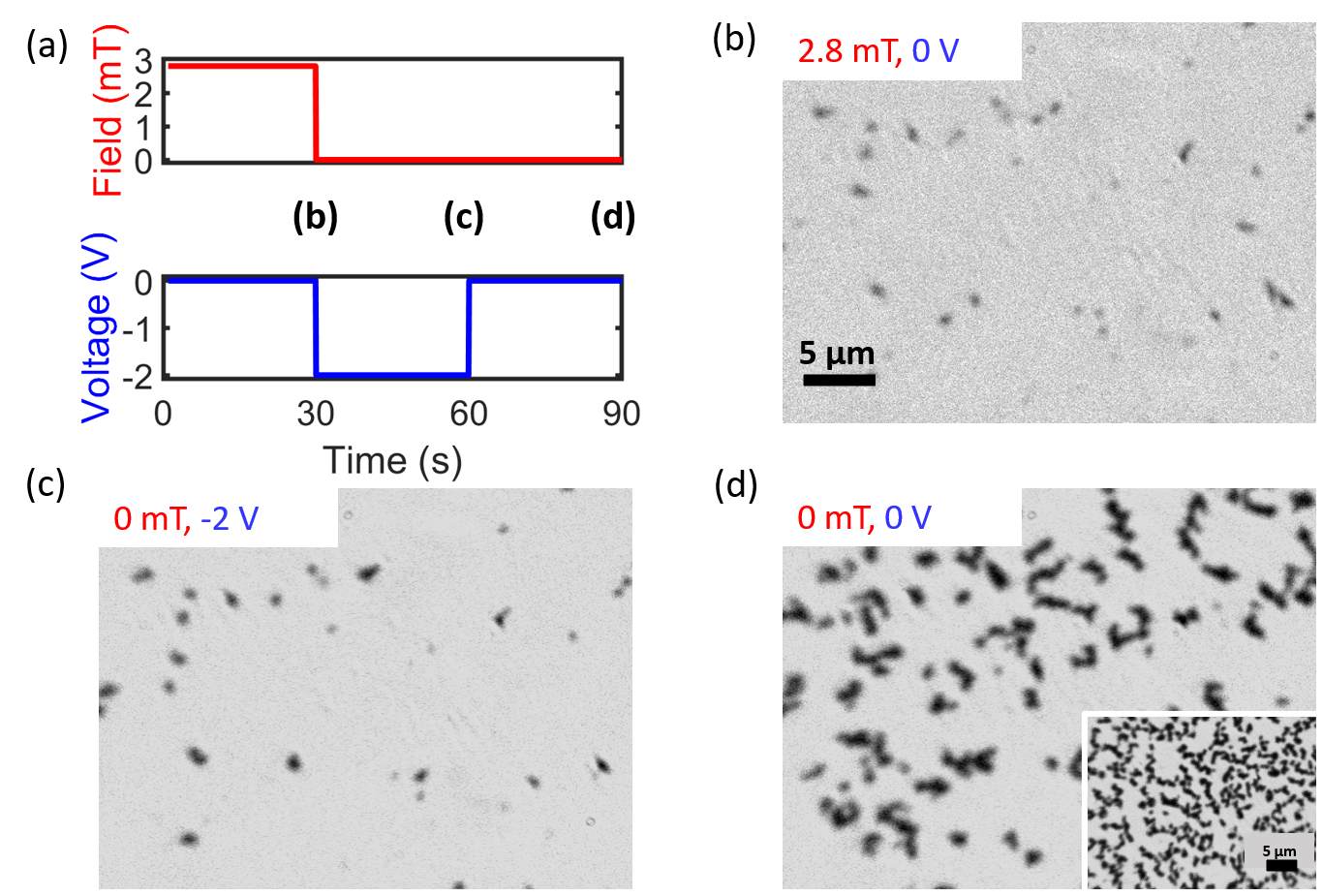}
\caption{(a) Schematic of applied magnetic field and voltage sequence. Times when the images in (b), (c) and (d) were captured are marked. (b) Skyrmion state at 2.8 mT and 0 V. (c) Skyrmion state at 0 mT and -2 V. (d) Multidomain state at 0 mT and 0 V. The inset shows the multidomain state at 0 mT that is attained directly from saturation at 10 mT with zero voltage.}
\label{fig3}
\end{figure}

Besides the on-demand creation and annihilation of skyrmions at 3.5 mT, we find that a negative bias voltage can stabilize skyrmions at zero magnetic field (Fig. 3). To demonstrate this, we first create a skyrmion state at 2.8 mT without a bias voltage (Fig.\ 3(b)).  Then we apply -2 V and the magnetic field is turned off immediately afterwards. After 30 s, the skyrmion state is still similar to the initial skyrmion state at 2.8 mT and 0 V (compare Fig.\ 3(c) and Fig.\ 3(b)). In order to confirm that it is the negative bias voltage that controls the stabilization of skyrmions in zero magnetic field, we turn off the voltage subsequently. A clear transition occurs as the skyrmions then expand and the sample shows a multidomain state, similar to that seen at zero magnetic field and voltage after saturation (Fig.\ 3(d)). Here, by increasing the PMA with a negative bias voltage, the skyrmions, once formed, are stabilized against expanding to form worm-like domains. 
\begin{figure}[!h]
\centering
\includegraphics[width=\columnwidth]{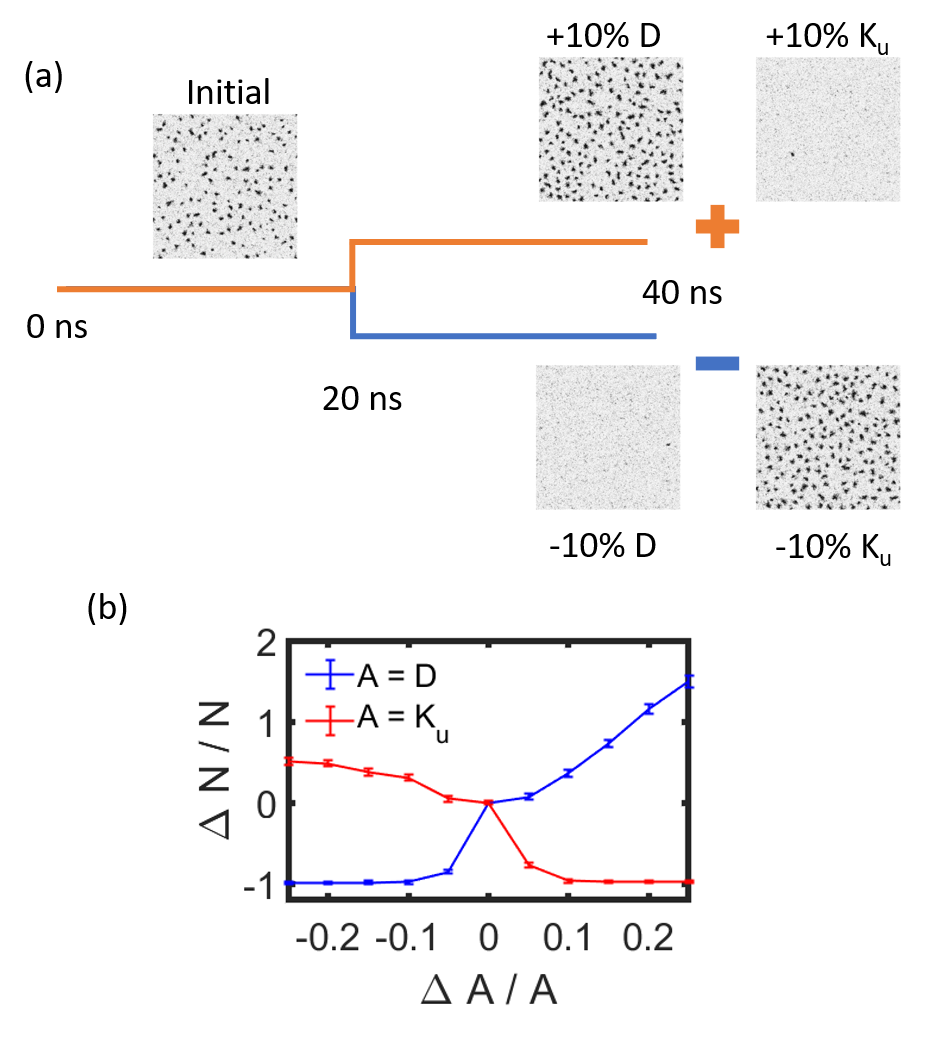}
\caption{(a) Simulated changes in the skyrmion state when parameters Ku and D are changed by +10\% and -10\% compared to their initial value. The orange and blue lines illustrate the timeline of the micromagnetic simulations. (b) The fractional change in skyrmion numbers due to a fractional change in $K_u$ or $D$.}
\label{fig4}
\end{figure}

Having shown voltage control of skyrmions in our system we turn to the question of underlying parameters being controlled by the applied voltage. It has been shown that voltages could influence $M_s$, $K_u$ and DMI $D$. We find that the saturated MOKE signals under different voltages are almost identical, indicating that $M_s$ remains the same. Due to a limited in-plane field in our MOKE system, we are not able to measure changes in $K_u$ directly. Instead, we perform micromagnetic simulations with the  MuMax3 package to gain insight into voltage effects on the skyrmion density. In order to achieve a spontaneous skyrmion state in simulations with a reasonable time scale, we adopt the following initial parameters:
$M_s = 0.8 \times 10^6$ A/m, exchange constant $A = 0.7 \times 10^{-12}$ J/m, $K_u = 0.5 \times 10^6$ J/m$^3$, $D = 1.5 \times 10^{-3}$ J/m$^{2}$ and damping constant $\alpha = 0.3$. A constant perpendicular magnetic field of 160 mT is applied to nucleate skyrmions, and the simulation temperature is set to 300 K.  Initially, the system is allowed to relax for 20 ns, and a snapshot of the magnetization is recorded. Then, either $K_u$ or $D$ are modified by a certain percentage of their initial value, and the resulting skyrmion state is recorded after a further 20 ns (Fig.\ 4(a)). The effect of changing $K_u$ or $D$ is directly illustrated by a change in the number of skyrmions, $\Delta N$, compared to the initial value ($\Delta N/N_0$). In Fig.\ 4(b), both an increase in $K_u$ and a decrease in $D$ lead to a decrease in $N$, and vice versa. Below 10 \% variation the effects of changing $K_u$ and $D$ are fairly symmetric, with changing $D$ being more effective for larger variation. 
By comparing simulation results to Fig.\ 1(c) and Fig.\ 2, we infer a negative bias voltage in our system either increases $K_u$, decreases $D$, or possibly both. 
A more quantitative comparison of simulations and our experimental results is not possible, due to the very different timescales involved.
\begin{figure}[!h]
\centering
\includegraphics[width=\columnwidth]{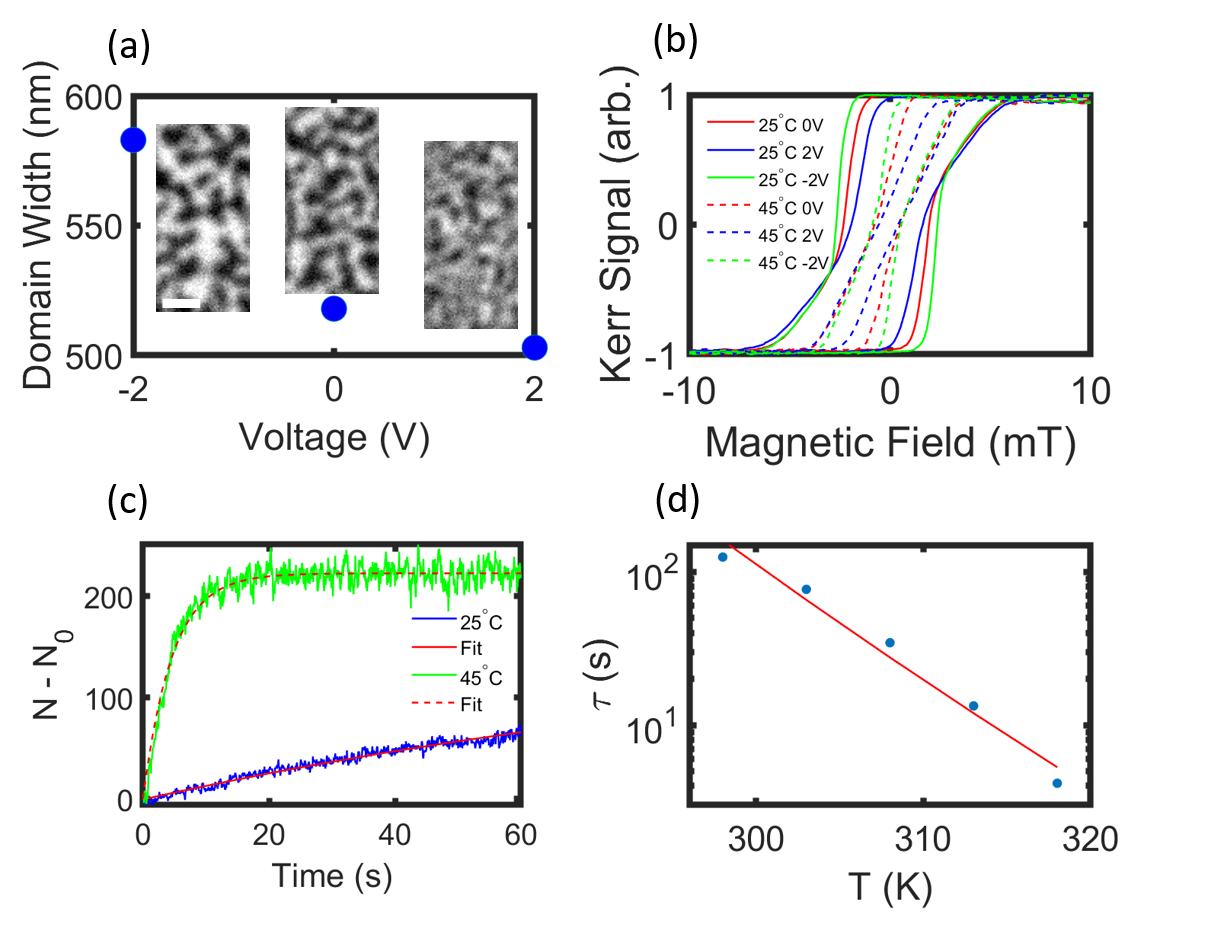}
\caption{(a) Average domain width obtained after demagnetization under applied 2 s voltage pulses of -2 V, 0 V and 2 V. The scale bar marks 5 $\mu$m.  (b) Hysteresis loops of the sample at 25$^{\circ}$C (solid lines) and 45$^{\circ}$C (dashed lines) with -2 V, 0 V and 2 V. (c) Time dependence of the variation of skyrmion numbers under 1.5 mT when applying a constant 2 V at 25$^{\circ}$C (blue) and 45$^{\circ}$C (green). (d) Fitting of the Arrhenius relation of the skyrmion creation time under 1.5 mT and 2 V at different temperatures.}
\label{fig5}
\end{figure}

Another open question is the underlying physical mechanism in our system. 
There are two main possible mechanisms of voltage control -- orbital filling effects or voltage-induced ion migration. 
In order to investigate this we first note that in our experiments, there are two relevant timescales: the relaxation time of the magnetic microstructure, which depends on the relevant thermal activation barrier; and the response time of the voltage-induced changes of magnetic parameters such as PMA and the DMI. For voltage effects driven by electron orbital filling we would expect the changes in the parameters to occur instantaneously on the timescales of our experiments. For ion migration driven changes the time scale is less clear, but could be expected to occur over a period of hours at room temperature.
Generally, even if the voltage controlled changes of magnetic parameters are fast, it takes more time for skyrmions to nucleate or annihilate because these processes are thermally activated. Distinguishing between the thermal-induced relaxation and the effect of voltages would allow us to elucidate the mechanism of voltage control. In order to study the timescale of the voltage effect, we first investigate the short time behavior of the sample under applied voltages. We demagnetized the sample by an AC oscillating field in 2 s in order to exclude magnetic relaxation effects in the Co layer. A voltage pulse is applied simultaneously with the demagnetization field and turned off immediately after the sample is demagnetized. Hereafter, an image of the zero-field domain state is taken immediately. 
From the domain states found for different voltages, the average domain width of the sample is extracted by a Gaussian fitting to the fast Fourier transformation of the domain images (Fig.\ 5 (a)). Clearly, 2 s voltage pulses affect the domain width in the demagnetized state, where a -2 V pulse increases the average domain width and 2 V has the reverse effect. This is consistent with the changes in the skyrmion density found with longer pulses, showing that significant effects are seen within 2 s of applying the voltage. 

To study how voltage-induced changes of the magnetic parameters affect the nucleation of skyrmions on longer time scales, we investigate the thermal activation of skyrmions by conducting experiments at different temperature.
In Fig.\ 5(b), voltage effects on the magnetic hysteresis loop are shown at 25$^{\circ}$C and 45$^{\circ}$C in a junction exhibiting a skyrmion state from $1$ mT to $2$ mT depending on the temperature. To investigate skyrmion nucleation we first relax the sample at $1.5$ mT, then apply a constant 2 V bias and count the number of skyrmions as a function of time for 1 min. Since a positive voltage could decreases the PMA or increase the DMI, more skyrmions are created following the application of the voltage through thermal activation (Fig.\ 5(c)). By assuming that the voltage effect is effectively instantaneous and does not change with time, the following time-dependent equation of skyrmion number $N$ can be written with a scale factor $A$, a constant skyrmion creation time $\tau$ and initial number of skyrmions $N_0$\cite{wilson2019measuring}:
\begin{equation}
N-N_0 = A(1-\exp[-t/\tau]).
\end{equation}
The data of $N$ as a function of $t$ is collected at different temperatures: 25$^{\circ}$C, 30$^{\circ}$C, 35$^{\circ}$C, 40$^{\circ}$C and 45$^{\circ}$C. The value of $\tau$ is fitted at each temperature. If the nucleation process is determined by a single energy barrier, then we would expect the rate to follow an  
Arrhenius law:

\begin{equation}
    \tau = \tau_{0} \exp[- \frac{E}{k_BT}],
\end{equation}
where $\tau_{0}$ is the scale factor, $E$ is the energy barrier, $k_B$ is the Boltzmann constant and $T$ is the temperature. 
As shown in Fig.\ 5(d), we fit $\tau$ to Eq. (2), which gives a nucleation energy barrier $E = 1.4$ eV for skyrmions at 2 V and 1.5 mT. 
Our assumption of a constant creation rate and single energy barrier is not explicitly violated by this data. Combined with our other experiments, this means that the voltage effects on the parameters determining the energy barrier are likely to be near instantaneous without longer-time effects. The direction of voltages effects, i.e. decreasing PMA with positive voltage and vice versa, as well as their volatility, is consistent with that found by ab-initio calculation of electronic effects at Fe/MgO interfaces, assuming that the first interfacial Fe layer is oxidized\cite{zhang2017model}. The voltage effect on skyrmions seen here also has the same sign as in the Co/AlO$_x$ system\cite{schott2017skyrmion}. Previously it has been reported that changes in a Pt/Co/GdO$_x$ system were driven by ion migration, where the mobile ions in the GdO$_x$ layer, were determined to largely come from the atmosphere\cite{tan2019magneto}. The apparent lack of ion migration in our system may originate from the reversed layer sequence, where, in our case, the GdO$_x$ layer is buried under the top electrode, blocking ion diffusion originating from the atmosphere.

In conclusion, we demonstrated on-demand creation and annihilation of skyrmions by an applied electric field in a GdOx/Gd/Co/Pt structure. Additionally, we developed a method to stabilize skyrmions in zero magnetic field by voltage. We have investigated through simulations the possible underlying magnetic parameters influenced by the voltages. The simulation results show that changes in PMA and DMI have similar effects on the skyrmion density, but with opposite signs. We also looked at the timescale of the effects, showing that the voltages have an effect within 2 s, without changing significantly on a longer timescale. We conclude that the voltage effects derive from the modification of the orbital filling at the Co/GdO$_x$ interface meaning that they are in principle limited by capacitive effects. Our results show voltage controlled skyrmion effects that could be exploited for device physics and should encourage further work in this field. 

\section*{Acknowledgements}
This work was supported by the Academy of Finland (Grant  Nos.  295269,  306978 and 327804). We acknowledge the provision of facilities by Aalto University at OtaNano, the Micronova Nanofabrication Center and the Nanomicroscopy Center, as well as computational resources provided by the Aalto Science-IT project.

The data that support the findings of this study are available from the corresponding author upon reasonable request.

\bibliographystyle{aipnum4-1}
\bibliography{Reference}
\end{document}